\documentclass[journal=jacsat,manuscript=article]{achemso}

\usepackage[version=3]{mhchem} % Formula subscripts using \ce{}
\usepackage{braket}
\usepackage{amssymb}
\usepackage{amsmath}
\usepackage{subcaption}
\usepackage{xcolor}
\usepackage{xr-hyper}
\usepackage[normalem]{ulem}
\author{Hanwen Jin}
\affiliation[First University]
{Department of Materials, Imperial College London, England}
\alsoaffiliation[CSRC]
{Material and Energy Division, Beijing Computational Science Research Centre, China}
\author{Chengcheng Xiao}
\affiliation[First University]
{Department of Materials, Imperial College London, England}
\author{Matias Herran}
\affiliation[Third University]
{Nanoinstitut München, Fakultät für Physik, Ludwig-Maximilians-Universität München, Munich, Germany}
\author{Emiliano Cort\'es}
\affiliation[Third University]
{Nanoinstitut München, Fakultät für Physik, Ludwig-Maximilians-Universität München, Munich, Germany}
\author{Shiwu Gao}
\affiliation[CSRC]
{Material and Energy Division, Beijing Computational Science Research Centre, China}
\author{Johannes Lischner}
\email{j.lischner@imperial.ac.uk}
\affiliation[First University]
{Department of Materials, Imperial College London, England}
\alsoaffiliation[Second University]
{The Thomas Young Centre for Theory and Simulation of Materials, London E1 4NS,
United Kingdom}
\title[An \textsf{achemso} demo]
  {Hot-carrier generation in bimetallic Janus nanoparticles}

\abbreviations{IR,NMR,UV}

\newcommand{\AuAgAgSideAngleEnhancementElectrons}{1.34}
\newcommand{\AuAgAgSideAngleEnhancementHoles}{1.39}
\newcommand{\AuAgAuSideAngleEnhancementElectrons}{1.25}
\newcommand{\AuAgAuSideAngleEnhancementHoles}{1.82}

\begin{document}

\begin{tocentry}
    \includegraphics[height=4.45cm]{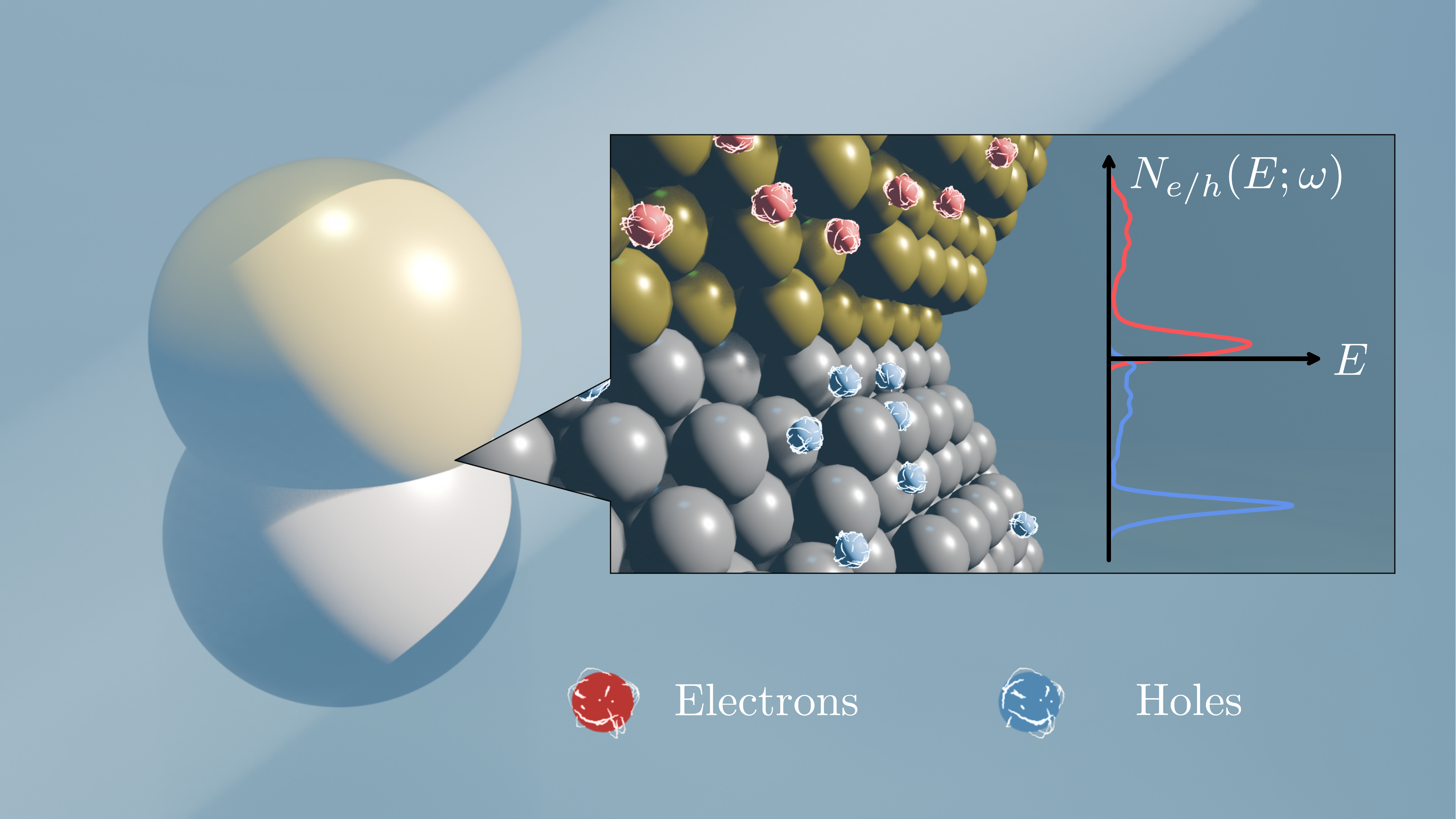}
\end{tocentry}


\begin{abstract}
  Energetic electrons and holes generated from the decay of localized surface plasmons in metallic nanoparticles can be harnessed in nanoscale devices for photocatalysis, photovoltaics or sensing. In this work, we study the generation of such hot carriers in bimetallic Janus nanoparticles composed of Au, Ag and Cu using a recently developed atomistic modelling approach that combines a solution of the macroscopic Maxwell equation with large-scale quantum-mechanical tight-binding models. We first analyze spherical Janus nanoparticles whose unique hot-carrier spectrum can be associated with the spectra of the two hemispheres and the interface coupling and find that under solar illumination the Ag-Au system exhibits the highest hot-carrier generation rate. For dumbbell-shaped Janus nanoparticles, we observe a significant increase in hot-carrier generation with increasing neck size. This is caused by a dramatic enhancement of the electric field in the neck region. We also study the dependence of hot-carrier generation on the light polarization and find that the largest generation rates are obtained when the electric field is perpendicular to the interface between the two metals due to the maximal dipole coupling with the electric field. The insights from our study will guide the experimental design of efficient hot-carrier devices based on bimetallic Janus nanoparticles. 
\end{abstract}

\section{Introduction}
Metallic nanoparticles of materials such as Au, Ag or Cu are highly efficient light absorbers in the visible spectrum due to localized surface plasmons (LSPs) - collective oscillations of conduction electrons that decay via Landau damping into energetic or "hot" carriers. These hot carriers can be harnessed in devices for photocatalysis~\cite{Photocatalysis,Photocatalysis1,Photocatalysis2,Photocatalysis3,Photocatalysis4,Photocatalysis5,Erhart2}, photovoltaics~\cite{Photovoltaics,Photovoltaics1}, and photodetection~\cite{Antosiewicz2015,Swearer2016,Photodetection,Photodetection1,Photodetection2,Photodetection3,Photodetection4,Photodetection5}. However, the development of efficient hot-carrier devices faces significant challenges as the dependence of hot-carrier properties on nanoparticle morphology, composition and environment remains ill-understood~\cite{Govorov3, Plasmon_limit,EmilianoNP}.

To address this knowledge gap, early theoretical work by Manjavacas and coworkers~\cite{Nortlander1} and also by Govorov et al.~\cite{Govorov1,Govorov2} employed free electron gas models to study hot-electron generation rates and steady-state hot-carrier distributions. Building on this, Chen et al.~\cite{GSW} and also Wu et al.~\cite{GSW2} developed models to explain plasmon-enhanced O\textsubscript{2} dissociation reactions~\cite{Linic1}. More recently, atomistic first-principles techniques were used to overcome limitations of simple free-electron gas models: for example, Bernardi et al.~\cite{Louie1} and also Sundararaman and coworkers~\cite{Sundararaman2014} performed ab initio calculations on bulk metals to provide insights into hot-carrier properties, while several groups employed time-dependent density functional theory (TDDFT) to investigate hot carriers in small metallic nanoparticles~\cite{Rossi1,Rossi2,Rossi3,Ag923}. To simulate larger nanoparticles of relevance to realistic devices, Jin et al. developed a multi-scale approach that combines large-scale tight-binding models\cite{Cristian1,Cristian2} with a solution of the macroscopic Maxwell equation~\cite{Hanwen2022}. Using this approach, Kang et al. were able to explain the experimentally observed dependence of CO\textsubscript{2} reduction efficiency on the shape of Au nanoparticles~\cite{SimaospaperNatComms}.

While most theoretical work thus far has focused on nanoparticles composed of a single material, bimetallic nanomaterials offer additional opportunities for tuning hot-carrier properties. For example, Ramachandran et al. demonstrated that in spherical Ag-Au alloy nanoparticles, the relative contributions of interband and intraband transitions to the total hot-carrier generation can be controlled through the Ag-Au ratio~\cite{Shreyaspaper}. Fojt and coworkers observed that surface-alloying of Ag nanoparticles with transition metals significantly enhances hot-hole generation in the surface layer~\cite{Erhart1}. Jin and coworkers analyzed hot-carrier generation in Pd-Au nanomaterials, finding substantially higher generation rates in antenna-satellite architectures\cite{Naomi1,Naomi2,Naomi3} compared to Au@Pd core-shell nanoparticles~\cite{Hanwen2023}. These findings are consistent with the experimental findings of Herran et al., who observed that Pd-Au antenna-satellite systems exhibit a much higher increase in H\textsubscript{2} production upon illumination than core-shell nanoparticles~\cite{HerranEmi,Naomi1,Naomi2,Naomi3}. 

%From an optical point of view, the advantage of this structures over other combinations rely on the fact that the incomplete coverage of the components enables to maintain the plasmonic properties of the components, avoiding an extreme dampening (that depends of course on the losses), whereas from a catalytic point of view, every single janus offers reactive sites of different nature. Hence, light driven chemical processes e.g. redox, could take place within the same catalyst. 

Janus nanoparticles are another class of bimetallic systems in which each hemisphere is composed of a different metal, see Fig.~\ref{fig:janus_nanoparticles}. These systems exhibit many attractive properties, including tuneable optical properties, charge separation at the interface as well as different reaction sites allowing reduction and oxidation processes to occur in the same catalyst, and are used for applications in catalysis, photonics and drug delivery~\cite{JanusCata,JanusCata1,JanusCata2,Tandom0,JanusDrug,JanusDrug1,watersplitting,watersplitting1}. For example, Cu/Ag and Cu/Au Janus nanoparticles exhibit a superior CO\textsubscript{2} reduction performance as they combine a low overpotential for CO\textsubscript{2} to CO reduction due to the Ag/Au component~\cite{Tandem1,Tandom2,Tandom3} with the strong C-C coupling of Cu~\cite{Hannes}. Moreover, Au/Ag Janus nanoparticles exhibit a 7-fold increase in oxygen reduction reaction activity compared to monometallic Ag nanoparticles, attributed to asymmetric charge transfer from Ag to Au and optimized oxygen binding at the bimetallic interface~\cite{AuAg_experiment2}. Finally, Ag/Au Janus nanoparticles have also been used as surface-enhanced Raman scattering sensors for the detection of toxic substances~\cite{AuAg_experiment3}. Despite the promise of Janus nanoparticles, not much is known about hot-carrier generation in these systems. For example, detailed knowledge of how hot-carrier properties depend on the composition, size and shape of the Janus nanoparticle is currently lacking.

In this paper, we investigate the properties of plasmon-induced hot carriers in Ag-Au, Ag-Cu and Au-Cu Janus nanoparticles using the atomistic method developed by Jin et al.~\cite{Hanwen2022}. We first analyze hot-carrier production in spherical Janus nanoparticles and find that under solar illumination the Ag-Au system generates the most highly energetic carriers. Next, we investigate dumbbell-shaped nanoparticles and find that hot-carrier generation is increased in these systems as a consequence of electric-field enhancement in the neck region. A particularly large increase is observed in the Ag-Au system and Au-Cu compared to the Ag-Cu systems. The insights from our findings will guide the experimental design of bimetallic Janus nanoparticles for efficient hot-carrier devices.

\section{Results and discussion}
\begin{figure}
    \centering
    \begin{subfigure}{0.49\textwidth}
         \centering
         \includegraphics[width=1.0\textwidth]{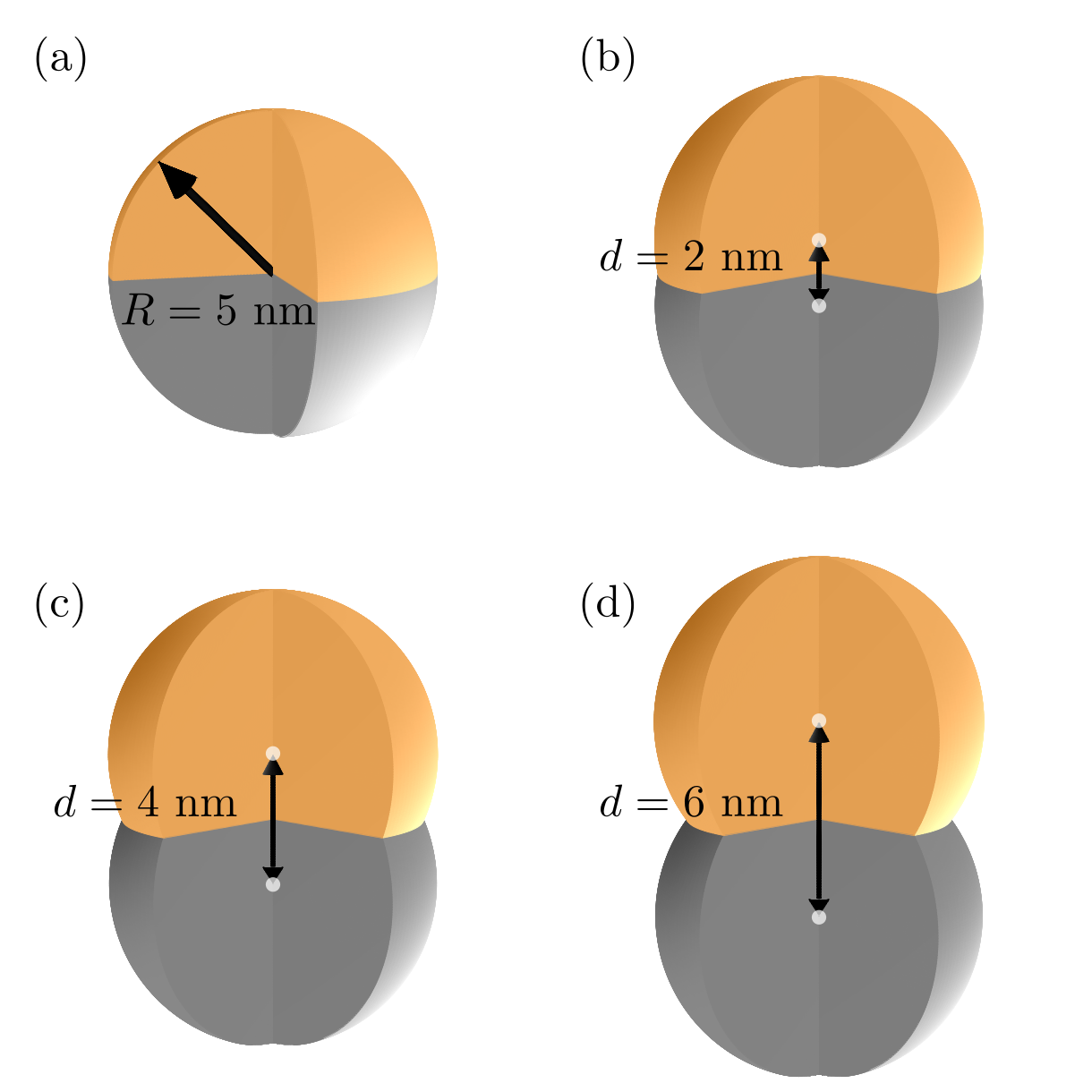}
     \end{subfigure}
     \begin{subfigure}{0.49\textwidth}
         \centering
         \includegraphics[width=1.0\textwidth]{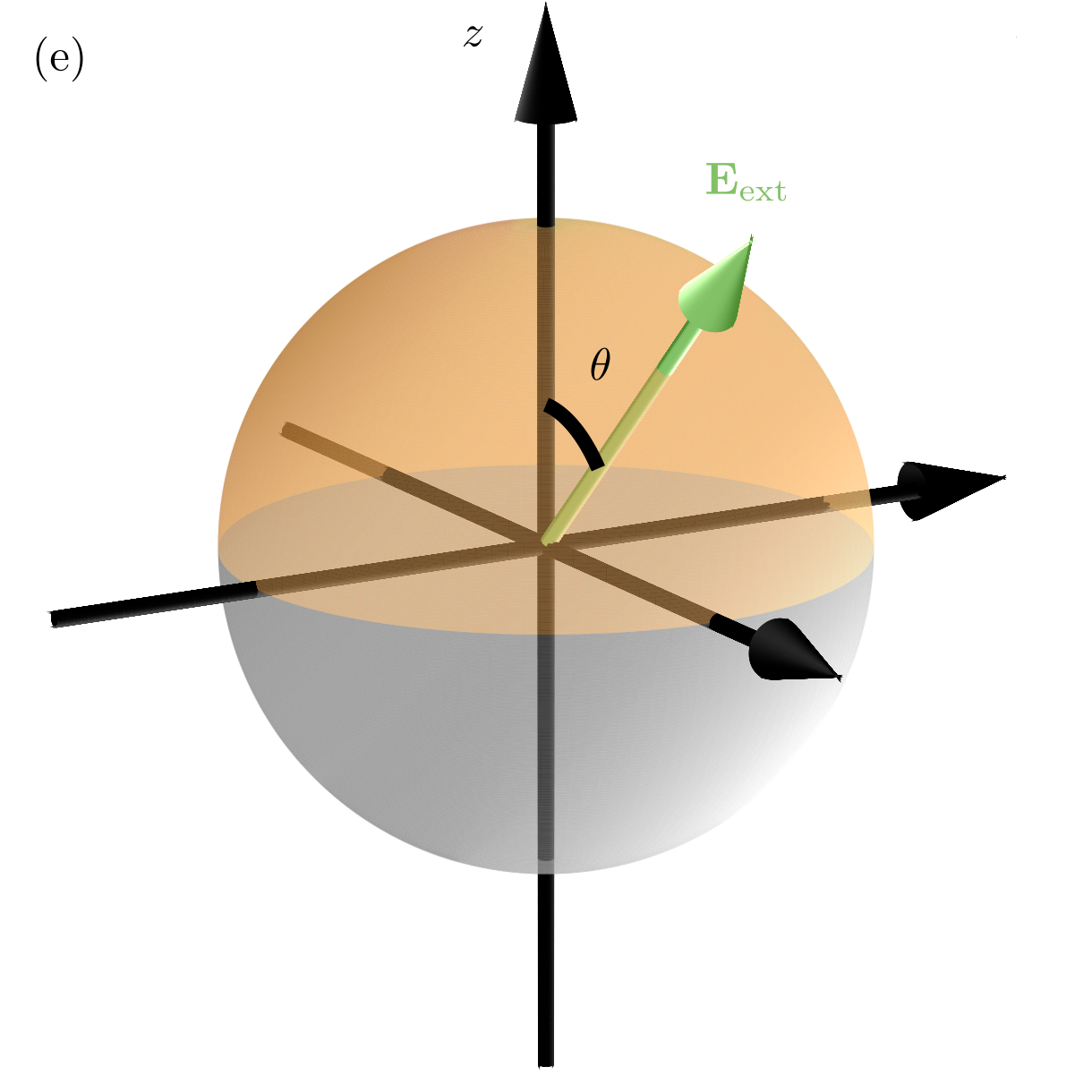}
     \end{subfigure}
    \caption{(a)-(d): Schematic illustration of the geometry of Janus nanoparticles studied in this work: a spherical Janus nanoparticle (a) and dumbbell-shaped Janus nanoparticles with different neck sizes $d$ (b)-(d). (e): Schematic illustrating the definition of the light polarization vector.}
    \label{fig:janus_nanoparticles}
\end{figure}

We study spherical Janus nanoparticles as well as dumbbell-shaped Janus nanoparticles, see Fig~\ref{fig:janus_nanoparticles}. The dumbbell-shaped nanoparticles are characterized by a neck size $d$ which is defined as the distance between the centers of the truncated spherical nanoparticles which are combined to form the dumbbell. The spherical Janus nanoparticles have a radius of 5 nm (containing 29,334 atoms for the Ag-Au system, 36,498 for the Au-Cu system and 36,498 atoms for the Ag-Cu system). For the dumbbell-shaped systems consisting of two truncated spheres, center-to-center distances, i.e. neck sizes, ranging from 2 nm to 6 nm (containing up to 65,456 atoms) are modelled. We define the z-axis as the normal to the interface between the two metals and the polarization angle $\theta$ as the angle between the z-axis and the external field $\mathbf{E}_{\text{ext}}$, see Fig.~\ref{fig:janus_nanoparticles}(e).

\subsection{Spherical Janus nanoparticles}

\begin{figure}
    \centering
    \includegraphics[]{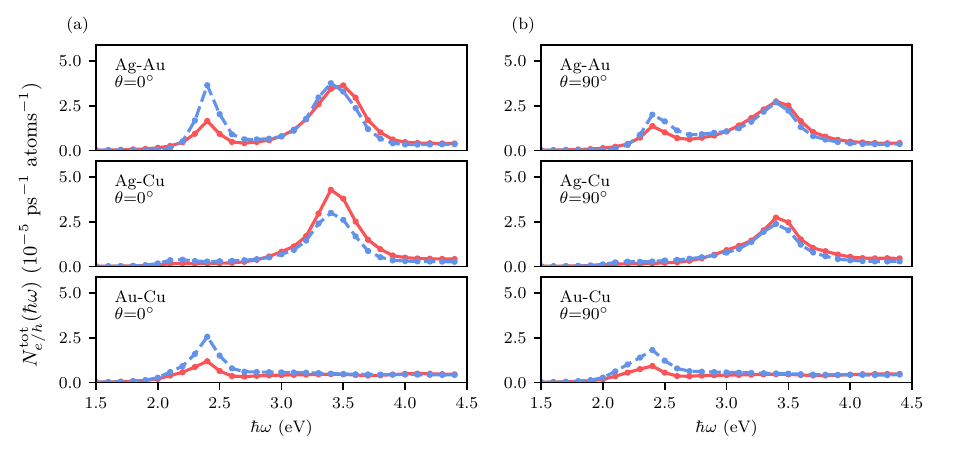}
    \caption{Total generation rate of highly energetic electrons (red solid line) and hot holes (blue dashed line) as function of photon energy for spherical Janus nanoparticles with different compositions (Ag-Au, Ag-Cu and Au-Cu). Highly energetic carriers are defined as having energies larger than 1 eV relative to the Fermi level. (a): Results for light polarization perpendicular to the interface ($\theta=0^\circ$). (b): Results for light polarization parallel to the interface ($\theta=90^\circ$).}
    \label{fig:freq_dep}
\end{figure}

Figure~\ref{fig:freq_dep} shows the total generation rate of highly energetic electrons and holes as function of the photon energy $\hbar \omega$ for Ag-Au, Ag-Cu and Au-Cu Janus nanoparticles (details of the calculations are described in the Methods section). Here, highly energetic carriers are defined as electrons (red solid lines) or holes (blue dashed lines) with energies more than $1$~eV from the Fermi level. For Ag-Au Janus nanoparticles, the generation rates of highly energetic electrons and holes exhibit two peaks: the broad peak near 3.4 eV corresponds to the LSP energy of Ag and the sharper peak near 2.4 eV to the LSP energy of Au. 

\begin{figure}
    \centering
    \includegraphics[]{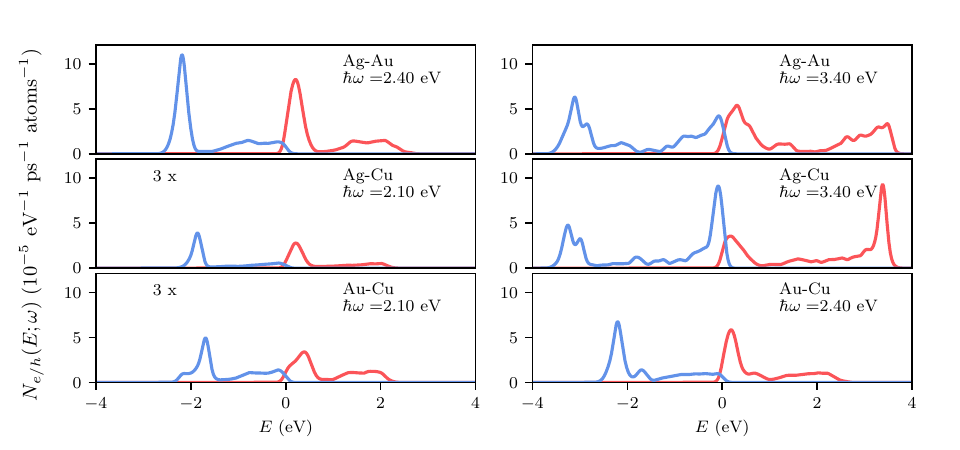}
    \caption{Energetic distribution of hot electrons (red lines) and hot holes (blue lines) in spherical Janus nanoparticles at the localized plasmon resonance frequencies, see Fig.~\ref{fig:freq_dep}. The electric field is perpendicular to the interface between the metals ($\theta=0^\circ$). For Janus nanoparticles containing Cu, the total generation rates of highly energetic electrons and holes only exhibit a single peak and we instead show the energetic distributions at the absorption onset of Cu at 2.1 eV. All energies are relative to the Fermi level.}
    \label{fig:janus_HCG}
\end{figure}

Whereas at a photon energy of 3.4 eV an equal amount of highly energetic electrons and holes are generated, at a lower photon energy of 2.4 eV the amount of holes is almost doubled in comparison to the electrons. To understand this finding, we analyze the energetic distribution of the excited electrons and holes at the two resonances, see Fig.~\ref{fig:janus_HCG}. At 2.4 eV, the hole distribution exhibits a large peak near -2 eV, while the corresponding peak in the electron distribution is close to the Fermi level. These peaks are caused by interband transitions in which electrons from Au d-states are excited into states with an sp-character. In addition to the interband peak, the electron and hole distributions also exhibit a second broader peak: for the holes, this peak is located close to the Fermi level, while for the electrons it is centered at approximately 2 eV. This peak arises from intraband transitions in which electrons transition from occupied states with sp-character to unoccupied states with the same character. Such intraband transition are forbidden by momentum conservation in the bulk material, but become possible in the nanoparticle. It can be seen that the intraband peak is significantly smaller than the interband peak explaining why a majority of highly energetic holes are generated at the 2.4 eV resonance. In contrast, the intraband peak is much stronger at the 3.4 eV resonance resulting in the approximately equal generation rate of highly energetic electrons and holes. The increased relevance of intraband transitions at the higher-energy resonance originates from the deeper-lying d-states in Ag making it more difficult to excite interband transitions.

Comparing the results for light polarizations parallel ($\theta=90^\circ$) and perpendicular ($\theta=0^\circ$) to the interface between the two metals, we find that more highly energetic carriers are generated when the electric field is perpendicular to the Ag-Au interface, see Fig.~\ref{fig:freq_dep}. Specifically, at the 3.4 eV resonance the generation rate of highly energetic electrons is enhanced by a factor of \AuAgAgSideAngleEnhancementElectrons, while the generation rate of highly energetic holes is enhanced by a factor of \AuAgAgSideAngleEnhancementHoles. At the 2.4 eV resonance, the enhancement is \AuAgAuSideAngleEnhancementElectrons ~for electrons and \AuAgAuSideAngleEnhancementHoles ~for holes. When the polarization is perpendicular to the interface ($\theta=0^\text{o}$), coupled plasmonic modes are excited across both Au and Ag regions, leading to field enhancement on both sides. In contrast, for parallel polarization ($\theta =90^{\text{o}}$), the plasmonic responses of the Ag and Au hemispheres remain largely decoupled and the field enhancement is primarily localized to the region with the dominant LSP resonance at the excitation frequency.

For the Ag-Cu system, the total generation rate of highly energetic carriers also exhibits a peak at 3.4 eV, see Fig.~\ref{fig:freq_dep}. Contrary to the Ag-Au system, more highly energetic electrons are produced at the resonance. This is a consequence of the dramatic increase in the production of hot electrons from intraband transitions, see Fig.~\ref{fig:janus_HCG}. Besides the 3.4 eV resonance, no additional peaks are observed in the generation rate of highly energetic carriers. This is consistent with our previous work~\cite{Hanwen2022} where we found that Cu nanoparticles in vacuum do not exhibit a clear plasmonic peak in their absorption spectrum. Again, we find that fewer highly energetic carriers are produced when the light polarization is parallel to the interface ($\theta=90^\circ$).

Finally, the generation rate of highly energetic carriers of the Au-Cu system exhibits a single peak at 2.4 eV. Similar to the Ag-Au system, more highly energetic holes are produced at this resonance. Comparing the three different Janus nanoparticles, we find that most highly energetic carriers are generated in the Ag-Au system, while the fewest are produced in the Au-Cu system. This is a consequence of the large electric field enhancement in the Ag hemisphere at the LSP resonance. However, the high energy of this resonance results in a small overlap with the solar spectrum.

\subsection{Dumbbell-shaped Janus nanoparticles}

\begin{figure}
    \centering
    \includegraphics[]{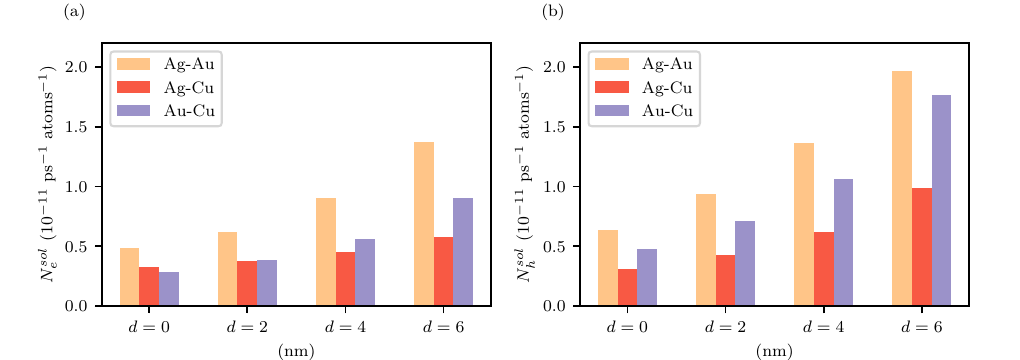}
    \caption{Total generation rate of (a) highly energetic electrons and (b) and highly energetic holes of Ag-Au, Ag-Cu and Au-Cu Janus nanoparticles as function of neck size $d$ under solar illumination. Highly energetic carriers are defined as having energies larger than 1 eV relative to the Fermi level. The electric field is perpendicular to the interface ($\theta=0^\circ$).}
    \label{fig:neck_depend2}
\end{figure}

Figure~\ref{fig:neck_depend2} compares the total generation rates of highly energetic electrons and holes of Ag-Au, Ag-Cu and Au-Cu Janus nanoparticles with different neck sizes under solar illumination. We find that the Ag-Au systems produce the most highly energetic electrons and holes. Moreover, this material combination exhibits a significant increase in the production of highly energetic carriers as the neck size increases. The Au-Cu system also exhibits a strong increase in hot-carrier production with increasing neck size and always produces more hot holes than the Ag-Cu system.

\begin{figure}
    \centering
    \includegraphics[width=\linewidth]{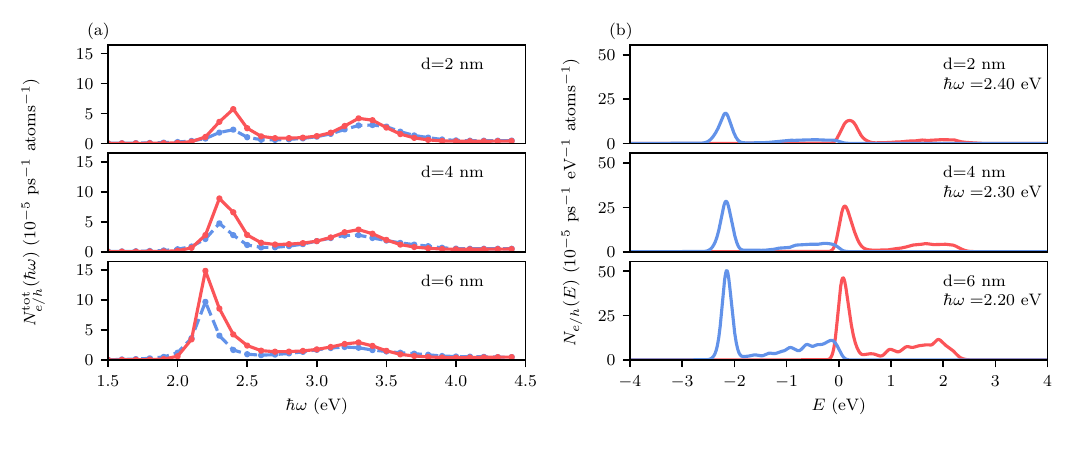}
    \caption{(a): Total generation rate of highly energetic electrons (red solid line) and holes (blue dashed line) as function of photon energy for Ag-Au Janus nanoparticles with different neck sizes. Highly energetic carriers are defined as having energies larger than 1 eV relative to the Fermi level. (b): Energetic distribution of hot electrons (red lines) and hot holes (blue lines) in Ag-Au Janus nanoparticles with different neck sizes at the lower-energy localized plasmon resonance frequencies. All energies are relative to the Fermi level. The light polarization is $\theta=0^\circ$. }
    \label{fig:AuAg_d_depend}
\end{figure}

To gain a deeper understanding of the Ag-Au systems with different neck sizes, Fig.~\ref{fig:AuAg_d_depend}(a) shows the total generation rate of highly energetic electrons and holes as function of photon energy. As the neck size increases, the lower-energy resonance red-shifts from 2.4 eV for $d=0$~nm to 2.2~eV for $d=6$~nm. Moreover, the height of the peak at 2.4 eV increases significantly. Specifically, the rate of highly energetic electrons produced at this photon energy increases almost threefold.

\begin{figure}
    \centering
    \includegraphics[scale=1]{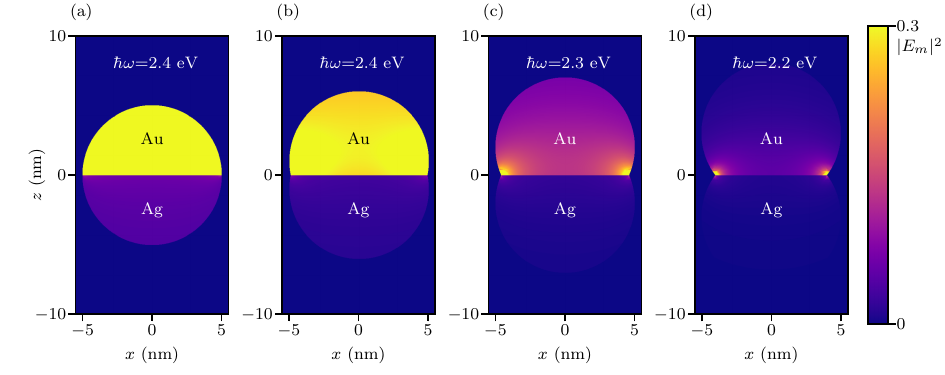}
    \caption{Squared magnitude of the electric field $|\mathbf{E}|^2$ (in units of the maximum squared field strength $|E_m|^2$) at the LSP energy of dumbbell-shaped Ag-Au nanoparticles with neck sizes (a) $d=0$ nm, (b) $d=2$ nm, (c) $d=4$ nm and (d) $d=6$ nm. For the $d=0$ nm, the squared maximum strength is $|E_m|^2 = 24 |E_0|^2$, for $d=2$~nm, it is $|E_m|^2 = 84 |E_0|^2$, for $d=4$~nm, it is $|E_m|^2 = 406 |E_0|^2$ and for $d=6$~nm, it is $|E_m|^2 = 3430 |E_0|^2$ with $E_0$ being the electric field strength of the external illumination. The electric field is perpendicular to the interface ($\theta=0^\circ$).}
    \label{fig:AuAg_field}
\end{figure}

The increase in hot-carrier generation is a consequence of spatial confinement of the electric field in the neck region, which enhances the local field strength, see Figure~\ref{fig:AuAg_field}. As the neck size increases from $d=0$~nm to $d=6$~nm, the maximum field enhancement $|E_m|^2$ grows dramatically from $24|E_0|^2$ to $3430|E_0|^2$, with intense field localization at the Ag-Au interface. The strong field enhancement at the gold plasmon frequency can be attributed to the frequency-dependent dielectric properties of the two metals. At 2.4 eV, Au exhibits moderate dielectric losses ($\epsilon = -3.21 + 1.86i$) while maintaining reasonable plasmonic response, enabling effective electromagnetic coupling with the neighboring Ag. In contrast, at the silver resonance regime (3.4 eV), gold exhibits significantly higher dielectric losses ($\epsilon = -0.66 + 5.49i$), which damps the electromagnetic coupling and reduces field enhancement.~\cite{dielectric_function} This frequency-dependent coupling behavior in bimetallic Ag-Au systems has been observed to produce enhanced plasmonic responses and significant electromagnetic field enhancement in the visible range. \cite{AuAg_experiment3}  Fig.~\ref{fig:AuAg_d_depend}(b) shows the energetic distribution of hot carriers at the low-energy resonance frequencies for the different neck sizes. It can be seen that the electric field enhancement results in a relatively uniform increase of the generation rates.

Finally, we discuss in more detail the Janus nanoparticles containing Cu. As Cu exhibits weaker plasmonic properties compared to Au and Ag, these systems produce fewer hot carriers. On the other hand, Cu is a better catalyst for many reactions, such as C-C coupling reactions. In Fig.~\ref{fig:neck_depend2}, we found that Au-Cu dumbbell-shaped nanoparticles produce more hot carriers than Ag-Cu ones. As the neck size increases in the Au-Cu system, the peak in the total generation rate of highly energetic carriers, shown in Fig.~\ref{fig:CuAu_d_depend}(a), increases significantly and also redshifts from 2.4 eV for $d=0$~nm to 2.1 eV for $d=6$~nm. The increase is caused by an additional contribution of hot carriers from interband transitions in the Cu component which produce an additional peak in the energetic distribution of hot holes at -1.5 eV, see Fig.~\ref{fig:CuAu_d_depend}(b). To understand the origin of this contribution, we analyze the electric field inside the Au-Cu nanoparticle, see Fig.~\ref{fig:CuAu_field}. At 2.1 eV, the electric field is distributed almost symmetrically in both the Cu and Au components indicating that a hybrid Au-Cu plasmon mode gets excited. The excitation of the Cu mode is possible because of the strong reduction of Cu's dielectric losses in this frequency range.

\section{Conclusion}
We have studied the generation of highly energetic electrons and holes resulting from the decay of localized surface plasmons in bimetallic Janus nanoparticles using an atomistic
modelling approach. In particular, we have performed calculations for Ag-Au, Ag-Cu and Au-Cu systems both for spherical as well as dumbbell-shaped nanoparticles. For Ag-Au and Au-Cu, we find a dramatic increase of the hot-carrier generation rates as the neck size increases while a smaller increase is observed for the Ag-Cu system. This increase is caused by the nanoscale confinement of the electric field in the neck region. Our finding opens up the exciting possibility of using Janus nanoparticles as photocatalysts in which the electric field exhibits a hot spot at the interface between the metals ensuring both high generation rates of hot carriers as well as offering multiple reaction sites and avoiding hot-carrier recombination.

\section{Methods}
\subsection{Atomic structure of Janus nanoparticles}

To construct atomistic models of Janus nanoparticles, we first create a flat interface between the two materials. For all combinations of materials (Ag/Au, Ag/Cu and Au/Cu), we focus on the (111) interface. As the lattice constants of Au and Ag are very similar, a $1 \times 1$ in-plane supercell is used whose lattice constant is the average of the Au and Ag lattice constants. For the Au/Cu and the Ag/Cu interfaces, the interface is created by combining a $7 \times 7$ Ag/Au in-plane supercell with an $8 \times 8$ Cu in-plane supercell, the supercells are shown in figure \ref{fig:cell}. The strain required to form a commensurate interface is less than 1 percent for each material. Once the flat interface has been created, we add atoms on both sides of the interface until the desired Janus nanoparticle shape is realized. 

\subsection{Tight-binding approach}
To construct tight-binding models for the Janus nanoparticles, we use a Slater-Koster approach \cite{Slater_Koster} retaining only nearest-neighbor and next-nearest neighbor hoppings. As in our previous work, we use a basis of 9 atomic orbitals per atom (e.g. for Au, the 5d, 6s and 6p orbitals are included)~\cite{Hanwen2022}.

To determine the tight-binding Hamiltonian parameters, we perform ab initio density-functional theory calculations using the Vienna Ab initio Simulation Package (VASP)~\cite{VASP1,VASP2}. We employ the Perdew-Burke-Ernzerhof (PBE) exchange-correlation functional~\cite{PBE} with a plane-wave cutoff energy of 500 eV. In these DFT calculations, we set the lattice constants to 4.15 \AA\ for gold and silver, and 3.63 \AA\ for copper. These values create a coherent lattice with induced strain less than 1\% compared to the fully relaxed structures, enabling coherent interfaces between the materials.

We compute Kohn-Sham band energies $E_{n\textbf{k}}^{\text{DFT}}$ on a $10\times 10 \times 10$ Monkhorst-Pack k-point mesh~\cite{MK_grid}. We obtain the optimal Slater-Koster parameters through least-squares fitting~\cite{NumericalRecipes} to minimize the deviation between DFT and tight-binding band structures. We take initial parameter values from Ref.~\cite{papa_book} and present the resulting band structure comparison in Fig.~\ref{fig:band}.

To obtain a tight-binding model for an interface between two materials, we use the tight-binding parameters of the bulk materials for the hopping matrix elements between atoms of the same type (e.g. Au to Au). For hopping matrix elements between different atomic species (e.g. Au to Ag), we simply average the bulk hopping matrix elements. To ensure a consistent choice of the onsite matrix elements of the two materials, we add a constant energy shift to the onsite matrix elements of each material. The value of this shift is found by comparing the density of states of the interface obtained from the tight-binding calculations to the result obtained from an ab initio DFT calculation for the same interface. The result for the fitted tight-binding model can be found in Table \ref{tab:SK}. The comparison to the DFT density of states can be found in Figure \ref{fig:dos}.

\subsection{Fermi's golden rule}
The hot-electron generation rates $N_e(E,\omega)$, i.e. the number of electrons generated with an energy $E$ when the nanoparticle is illuminated by light with frequency $\omega$, are obtained using a recently developed approach to efficiently evaluate Fermi's golden rule~\cite{Hanwen2022}
\begin{equation}\label{eq:Fermi_golden}
    N_e(E,\omega)=\frac{4\pi}{\hbar}\sum_{if}|\bra{i}\hat{\Phi}_{tot}(\omega)\ket{f}|^2\delta(E_i-E_f+\hbar\omega;\gamma)\delta(E-E_f;\sigma)f(E_i)(1-f(E_f)).
\end{equation}
Here $i$ and $f$ label the initial and final state, $\delta(x;\sigma)$ is a Gaussian broadened Dirac-delta function with a broadening of $\sigma$. We use broadenings of $\sigma=0.06$ eV to account for the electron lifetime and $\gamma=0.1$ eV to capture the transition linewidth. Finally, $f(E)=1/(1+\exp\left(E/k_BT\right))$ is the Fermi-Dirac distribution at $T=300$ Kelvin and $\hat{\Phi}_{tot}(\omega)$ is the quasi-static potential obtained by COMSOL Multiphysics\textsuperscript{\textregistered}\cite{quasi_static,Hanwen2023,comsol_multiphysics}, we used an external field strength of $E_0=7.53\times10^5$, corresponding to the illumination intensity of $10^9$ Wm$^{-2}$.  We used experimental measured dielectric functions from Ref.~\cite{dielectric_function} The dielectric constant of medium is 1.77, corresponding to the optical dielectric constant of water.

Naively, evaluating Eq.~\ref{eq:Fermi_golden} requires diagonalizing the electronic Hamiltonian of the nanoparticle. This scales cubically with the number of atoms and is not feasible for large nanoparticles. Instead, we use the kernel polynomial method~\cite{KPM,KITE} developed by Lischner et. al.~\cite{Hanwen2022} to evaluate the hot-carrier generation rate. This approach does not require the diagonalization of the Hamiltonian and scales linearly with the number of atoms in the nanoparticle.

We also calculate the generation rate of electrons and holes with energies larger than $E_{\text{thr}}=1$~eV from the Fermi level. These are given by 
\begin{equation}
    N_{e}^{tot}(\omega)=\int_{E_{\text{thr}}}^{\infty} N_e(E,\omega) dE. 
\end{equation}
and
\begin{equation}
    N_{h}^{tot}(\omega)=\int^{-E_{\text{thr}}}_{-\infty} N_h(E,\omega) dE. 
\end{equation}

Finally, we calculate the generation rates of highly energetic electrons and holes when the nanoparticles are exposed to solar illumination according to
\begin{equation}
    N_{e/h}^{sol}=\int_{0}^\infty  N_{e/h}^{tot}(\omega)S(\omega) d\omega,
\end{equation}
where $S(\omega)$ denotes the AM1.5 solar spectrum obtained from National Institute Standards and Institute (NIST) website. \cite{NIST}

\begin{acknowledgement}
S. Gao and H. Jin acknowledge support by the National Natural Science
Foundation of China (12393831, and 11934003). M. Harren and E. Cort\'es acknowledge the Deutsche Forschungsgemeinschaft (DFG) under Germany´s Excellence Strategy – EXC 2089/1 – 390776260, the Bavarian program Solar Technologies Go Hybrid (SolTech) and the Center for NanoScience (CeNS). J. Lischner acknowledges funding from the Royal Society through a Royal Society University Research Fellowship URF/R/191004 and also from the EPSRC programme grant EP/W017075/1.
\end{acknowledgement}

\bibliography{achemso-demo}

\end{document}

\begin{figure}
    \centering
    \includegraphics[width=\linewidth]{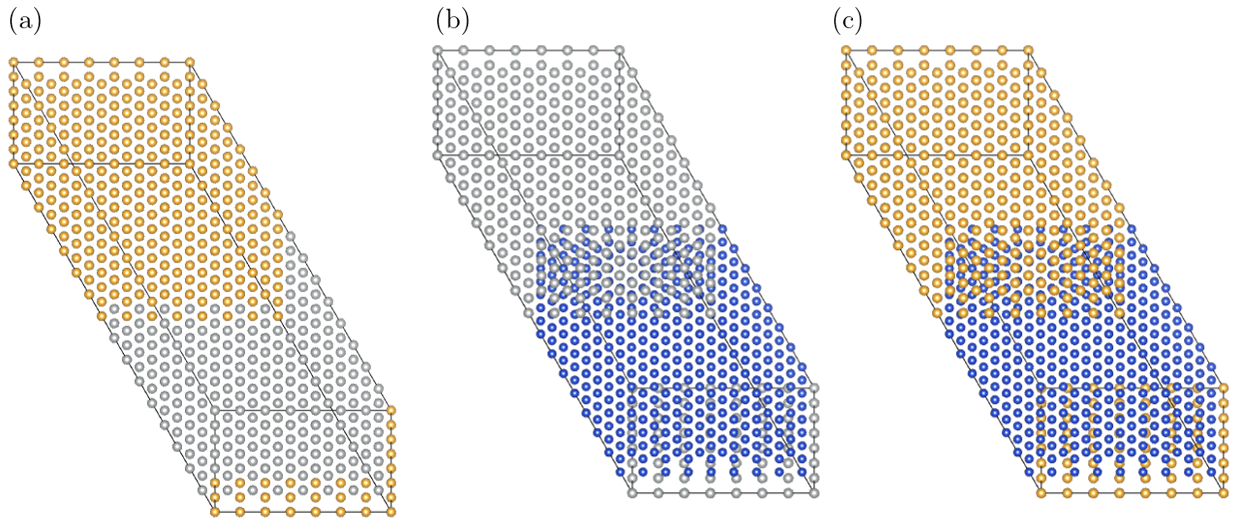}
    \caption{Atomic structure of the supercells used in the DFT calculations of the bimetallic interfaces. (a) Ag-Au interface, (b) Ag-Cu interface, (c) Au-Cu interface. Silver atoms are coloured in silver, gold atoms in gold and copper atoms in blue.  }
    \label{fig:cell}
\end{figure}

\begin{figure}
    \centering
    \includegraphics[]{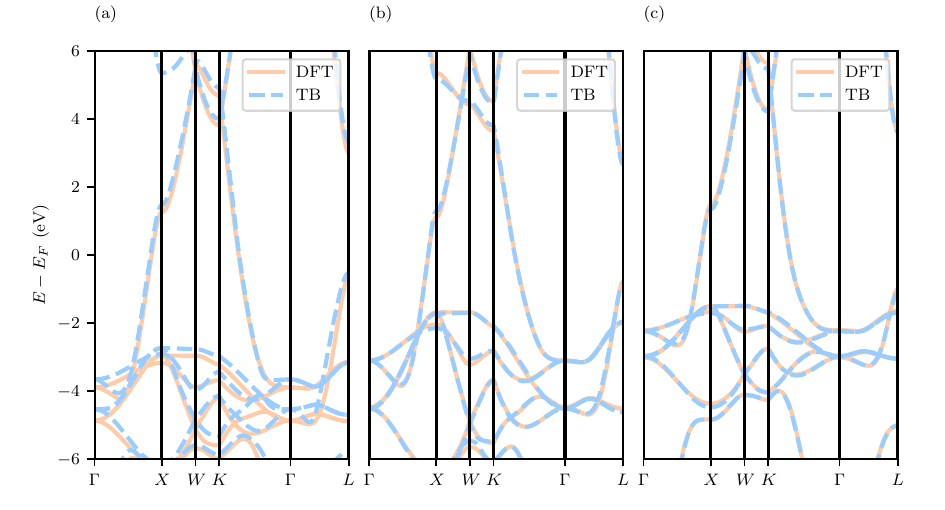}
    \caption{Bulk band structures of (a) Ag (b) Au (c) Cu calculated from DFT and tight binding. }
    \label{fig:band}
\end{figure}

\begin{figure}
    \centering
    \includegraphics[width=1\linewidth]{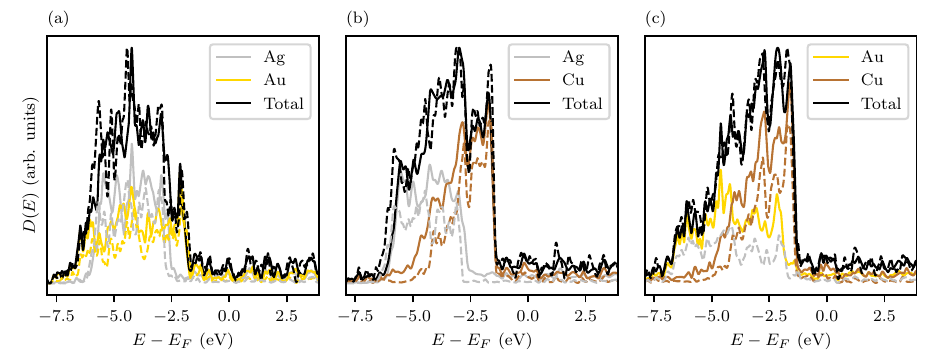}
    \caption{Density of states of (a) the Au/Ag interface, (b) the Cu/Ag interface, (c) the Cu/Au interface from tight-binding (solid lines) and the density functional theory (dashed lines). We have set the zero of the x-axis to the Fermi level of the system. }
    \label{fig:dos}
\end{figure}

\begin{table}[]
    \centering
    \caption{The Slater-Koster parameters (in eV) of Ag, Au and Cu. $\Delta E_{\text{Ag/Au/Cu}}$ denote the onsite energy shifts which are added to all onsite energies to ensure an accurate energy level alignment when a bimetallic interface is considered.}
    \label{tab:SK}
    \begin{tabular}{|c|c|c|c|}\hline
    &Ag&Au&Cu\\\hline
       $ss\sigma1$& -0.82304 & -0.88415 & -1.03045\\
        $pp\sigma1$& 2.10945 & 2.05236 & 2.25499\\
        $pp\pi1$ & -0.05893 & 0.01567 & 0.07597\\
        $dd\sigma1$ & -0.41541 & -0.63442 & -0.35511\\
         $dd\pi1$& 0.26730 & 0.28405 & 0.23146\\
         $dd\delta1$& -0.06275 & -0.02080 & -0.04289\\
         $sp\sigma1$& 1.17440 & 1.31924 & 1.45484\\
         $sd\sigma1$& 0.04737 & -0.65608 & -0.48483\\
         $pd\sigma1$& -0.03112 & -0.92976 & -0.56336\\
         $pd\pi1$& 0.49614 & 0.18467 & 0.19503\\
         $ss\sigma2$& -0.10597 & 0.03318 & 0.03886\\
        $pp\sigma2$& 0.19297 & 0.44646 & 0.58136\\
        $pp\pi2$ & 0.12400 & 0.05051 & 0.05634\\
        $dd\sigma2$ & -0.04779 & -0.03163 & -0.05987\\
         $dd\pi2$& 0.04316 & 0.00435 & 0.02339\\
         $dd\delta2$& 0.00738 & -0.00639 & -0.00577\\
         $sp\sigma2$& 0.01595 & -0.06003 & 0.08758\\
         $sd\sigma2$& 0.51901 & -0.09838 & -0.13260\\
         $pd\sigma2$& 0.04368 & -0.17655 & -0.11278\\
         $pd\pi2$& 0.32017 & 0.00579 & 0.01895\\
         $E_s$& 10.52798 & 9.55722 & 10.17554\\
         $E_p$& 16.86127 & 17.14594 & 16.43749\\
         $E_d$& 3.59990 & 5.09343 & 4.75432\\
         $a_0$& 4.15000 & 4.15000 & 3.63125\\
         $\Delta E_{\text{Ag}}$&0.0000&-0.60000&-0.10000\\
         $\Delta E_{\text{Au}}$&0.55000&-0.00000&-0.70000\\
         $\Delta E_{\text{Cu}}$&0.2000&0.80000&0.00000\\\hline
    \end{tabular}
\end{table}

%\begin{table}[]
%    \centering
%    \caption{Numerical values of the dielectric functions of bulk Ag, Au and Cu.}
%    \label{tab:epsw}
%    \begin{tabular}{|c|c|c|c|}\hline 
%    $\hbar\omega$ (eV)&Ag&Au&Cu\\\hline
%2.00&-17.40+2.26j&-9.97+0.82j&-10.42+1.75j\\
%2.10&-15.82+2.10j&-8.03+1.02j&-7.68+2.64j\\
%2.20&-14.24+1.95j&-6.39+1.22j&-6.07+4.32j\\
%2.30&-12.65+1.79j&-4.80+1.54j&-5.63+5.39j\\
%2.40&-11.07+1.64j&-3.21+1.86j&-5.51+5.82j\\\hline
%3.00&-5.10+1.04j&-0.87+5.54j&-3.49+5.22j\\
%3.10&-4.42+0.97j&-0.90+5.57j&-3.13+5.15j\\
%3.20&-3.75+0.89j&-0.87+5.54j&-2.77+5.09j\\
%3.30&-3.12+0.80j&-0.77+5.52j&-2.48+5.02j\\
%3.40&-2.55+0.70j&-0.66+5.49j&-2.19+4.95j\\
%\hline
%    \end{tabular}
%\end{table}

\begin{figure}
    \centering
    \includegraphics[width=\linewidth]{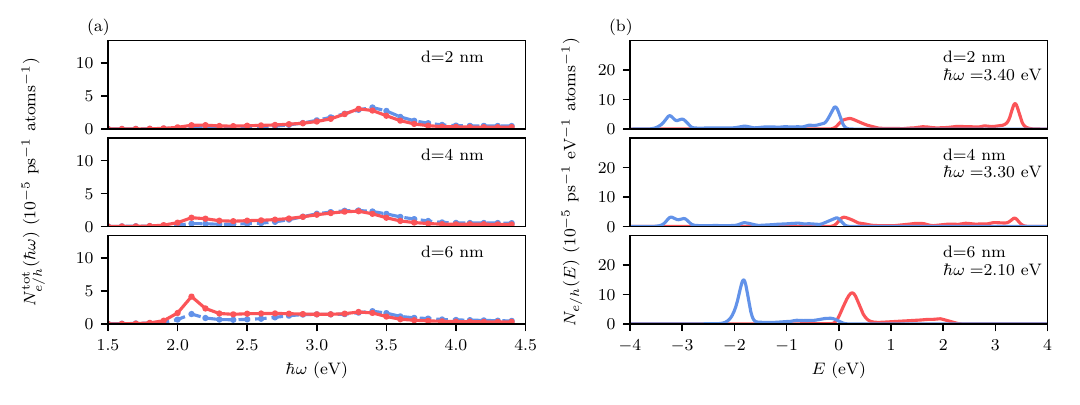}
    \caption{(a): Total generation rate of highly energetic electrons (red solid line) and holes (blue dashed line) as function of photon energy for Ag-Cu Janus nanoparticles with different neck sizes. Highly energetic carriers are defined as having energies larger than 1 eV relative to the Fermi level. (b): Energetic distribution of hot electrons (red lines) and hot holes (blue lines) in Ag-Cu Janus nanoparticles with different neck sizes at the lower-energy localized plasmon resonance frequencies. All energies are relative to the Fermi level. We used polarisation at $\theta=0^\text{o}$.}
    \label{fig:CuAg_d_depend}
\end{figure}

\begin{figure}
    \centering
    \includegraphics[width=\linewidth]{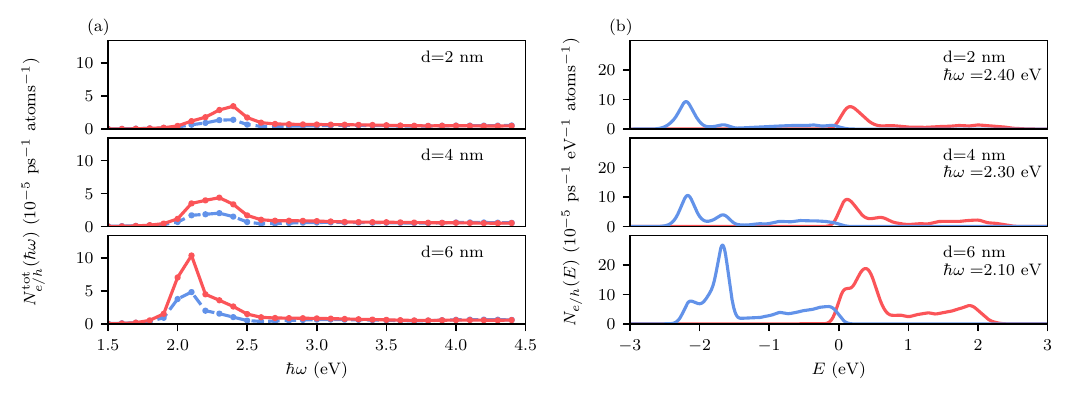}
    \caption{(a): Total generation rate of highly energetic electrons (red solid line) and holes (blue dashed line) as function of photon energy for Au-Cu Janus nanoparticles with different neck sizes. Highly energetic carriers are defined as having energies larger than 1 eV relative to the Fermi level. (b): Energetic distribution of hot electrons (red lines) and hot holes (blue lines) in Au-Cu Janus nanoparticles with different neck sizes at the lower-energy localized plasmon resonance frequencies. All energies are relative to the Fermi level. We used polarisation at $\theta=0^\text{o}$.}
    \label{fig:CuAu_d_depend}
\end{figure}

\begin{figure}
    \centering
    \includegraphics[]{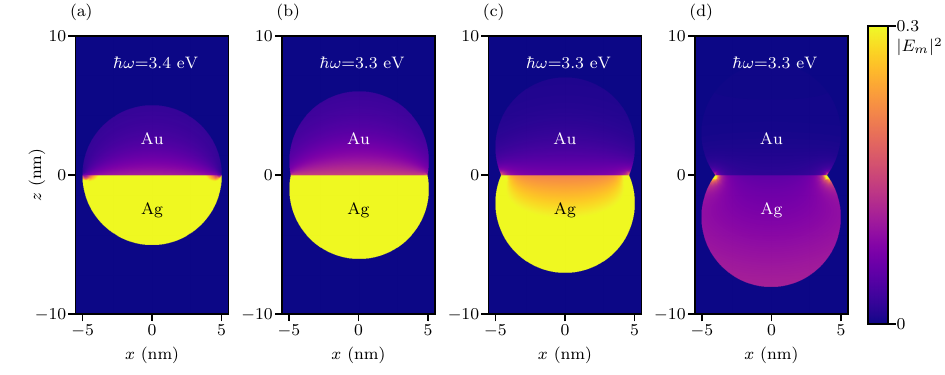}
    \caption{Electric field squared $|\mathbf{E}|^2$ (in unit of $E_0$) inside the four dumbell shaped Ag-Au nanoparticle considered in this thesis for the Ag plasmon (a) $d=0$ nm, with maximum intensity of 77, (b) $d=2$ nm, with maximum intensity of 73, (c) $d=4$ nm, with maximum intensity of 129 and (d) $d=6$ nm, with maximum intensity of 364, the polarisation is set at $\theta=0$. For better visualization purpose, have set the maximum value of in the color plot to be 0.3 times the absolute maximum value. }
    \label{fig:AuAg_Ag_field}
\end{figure}

\begin{figure}
    \centering
    \includegraphics[]{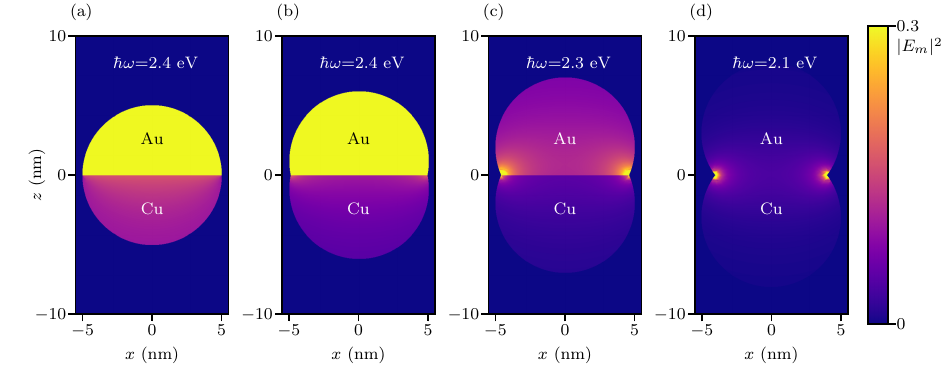}
    \caption{Electric field squared $|\mathbf{E}|^2$ (in unit of $E_0$) inside the four dumbell shaped Au-Cu nanoparticle considered in this thesis (a) $d=0$ nm, with maximum intensity of 15.7, (b) $d=2$ nm, with maximum intensity of 36.9, (c) $d=4$ nm, with maximum intensity of 191 and (d) $d=6$ nm, with maximum intensity of 1630, the polarisation is set at $\theta=0$. For better visualization purpose, have set the maximum value of in the color plot to be 0.3 times the absolute maximum value. }
    \label{fig:CuAu_field}
\end{figure}

\begin{figure}
    \centering
    \includegraphics[]{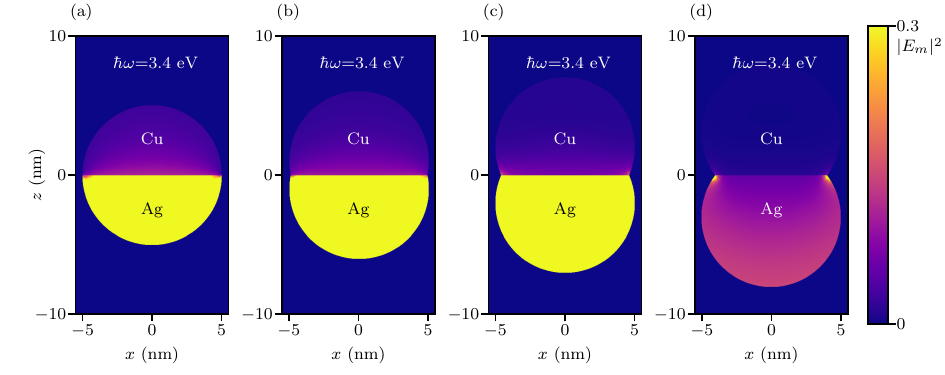}
    \caption{Electric field squared $|\mathbf{E}|^2$ (in unit of $E_0$) inside the four dumbell shaped Ag-Cu nanoparticle considered in this thesis (a) $d=0$ nm, with maximum intensity of 78, (b) $d=2$ nm, with maximum intensity of 75, (c) $d=4$ nm, with maximum intensity of 73 and (d) $d=6$ nm, with maximum intensity of 245, the polarisation is set at $\theta=0$. For better visualization purpose, have set the maximum value of in the color plot to be 0.3 times the absolute maximum value. }
    \label{fig:CuAg_field}
\end{figure}

\begin{figure}
    \centering
    \includegraphics[width=\linewidth]{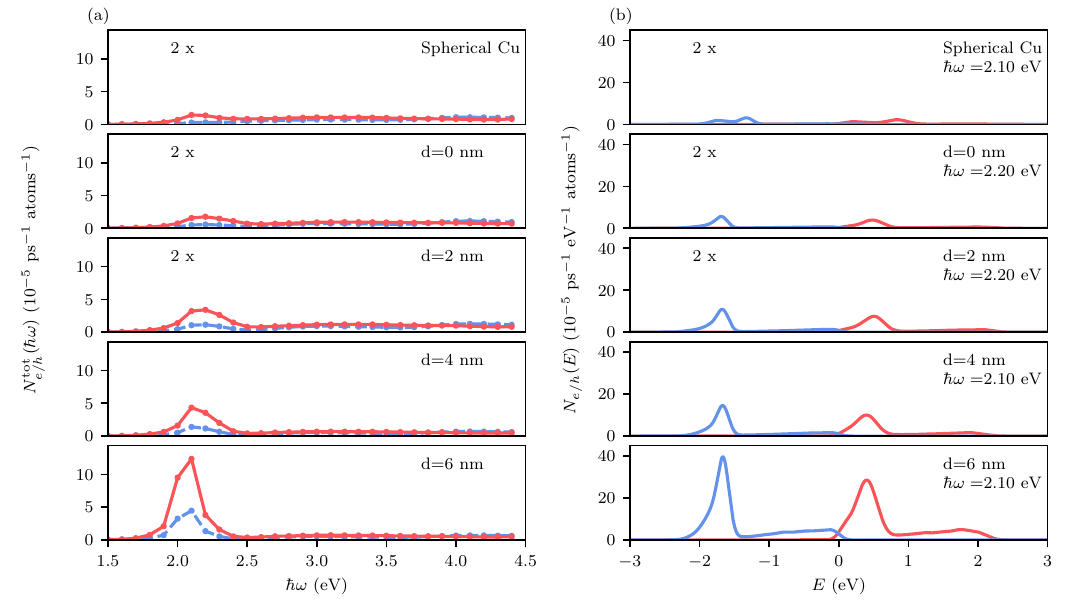}
    \caption{(a): Total generation rate of highly energetic electrons (red solid line) and holes (blue dashed line) as function of photon energy for the Copper part of Au-Cu Janus nanoparticles with different neck sizes, the top panel corresponds to spherical Cu nanoparticle as a reference. Highly energetic carriers are defined as having energies larger than 1 eV relative to the Fermi level. We separated the Cu contribution by treating Au as a dielectric, which is present in COMSOL multiphysics\textsuperscript{\textregistered} but absent in tight-binding. (b): Energetic distribution of hot electrons (red lines) and hot holes (blue lines) in Cu of Au-Cu Janus nanoparticles with different neck sizes at the lower-energy localized plasmon resonance frequencies. All energies are relative to the Fermi level. We used polarisation at $\theta=0^\text{o}$.}
    \label{fig:CuAu_Cu_d_depend}
\end{figure}